\PassOptionsToPackage{main=british}{babel}
\documentclass{moriond}

\usepackage[T1]{fontenc}
\usepackage[utf8]{inputenc}
\usepackage{amsmath}
\usepackage{amssymb}
\usepackage{microtype}
\usepackage{booktabs}
\usepackage{xfrac}
\usepackage{mleftright}
\usepackage{siunitx}
\usepackage{csquotes}
\usepackage[noabbrev]{cleveref}

\DeclareMathOperator{\Tr}{Tr}

\newcommand*{\lr}{\mleft(}
\newcommand*{\rr}{\mright)}

\newcommand*{\SUIIIC}{\operatorname{SU}(3)_{\text{C}}}
\newcommand*{\SUIIL}{\operatorname{SU}(2)_{\text{L}}}
\newcommand*{\UIY}{\operatorname{U}(1)_{\text{Y}}}

\newcommand*{\OmegacObs}{\Omega_{\text{c}}^{\text{obs}}}

\hypersetup{
	pdftitle={Singlet--doublet/triplet dark matter and neutrino masses},
	pdfauthor={Juri Fiaschi, Michael Klasen, Simon May},
	pdfkeywords={Beyond Standard Model, cosmology of theories beyond the SM, neutrino physics, discrete symmetries}
}

\newcommand*{\Photo}{}

\begin{document}
\vspace*{4cm}
\title{Singlet--doublet/triplet dark matter and neutrino masses}

\author{Juri Fiaschi$^*$, Michael Klasen$^*$ and \textbf{Simon May}$^{*,\dagger}$ \footnote{Speaker}}

\address{$^*$ Institut für Theoretische Physik, Westfälische Wilhelms-Universität Münster,\\
	Wilhelm-Klemm-Straße~9, 48149~Münster, Germany}

\address{$^\dagger$ Max-Planck-Institut für Astrophysik,\\
	Karl-Schwarzschild-Straße~1, 85741~Garching, Germany}

\maketitle\abstracts{
  In these proceedings, we present a study of a combined singlet--doublet fermion and
  triplet scalar model for dark matter (DM).
  Together, these models form a simple
  extension of the Standard Model (SM) that can account for DM and
  explain the existence of neutrino masses, which are generated radiatively.
  However, this also implies the existence of lepton flavour violating (LFV)
  processes. In addition, this particular model allows for gauge coupling
  unification. The new fields are odd under a new $\mathbb{Z}_2$ symmetry
  to stabilise the DM candidate. We analyse the DM,
  neutrino mass and LFV aspects,
  exploring the viable parameter space of the model. This is done using
  a numerical random scan imposing successively the neutrino
  mass and mixing, relic density, Higgs mass, direct detection, collider and
  LFV constraints. We find that DM in this model is
  fermionic for masses below about \SI{1}{\TeV} and scalar above.
  We observe a high degree of
  complementarity between direct detection and LFV
  experiments, which should soon allow to fully probe the fermionic DM sector and at least partially the scalar DM sector.%
}

\section{Introduction}

Particularly well-motivated DM models do not only provide a DM candidate,
but also solve other SM problems such as the smallness of neutrino masses.
This is possible when the $d = 5$ Weinberg operator is realised at one loop,\cite{Bonnet:2012kz} such that the particles in the loop have opposite $\mathbb{Z}_2$
parity to the SM particles and include a neutral DM candidate.\cite{Restrepo:2013aga}

In our paper,\cite{Fiaschi:2018rky} we
study a model of topology T1-3 with one scalar and two fermions, one of
which is vector-like. In contrast to the first of these models (T1-3-A
with hypercharge parameter $\alpha = 0$), where the scalar DM was a
singlet,\cite{Klasen:2016vgl} we investigate here a model (T1-3-B, also with
$\alpha = 0$) where the scalar DM is the neutral component of a triplet.
Both models,
like a previously studied model with both singlet--doublet scalars and
fermions (T1-2-A with $\alpha = 0$),\cite{Esch:2018ccs} have the additional
advantage that they allow for gauge coupling unification at a scale
$\Lambda = \mathcal{O}(\SI{e13}{\GeV})$.\cite{Hagedorn:2016dze}

\section{Description of the model}
\label{sec:model}

\begin{table}
	\centering
	\caption[]{New fields and their quantum numbers in the model T1-3-B with $\alpha = 0$.}
	\label{tab:model_content}
	\medskip
	\begin{tabular}{c c c c c c c c}
		\toprule
		Field & Generations & Spin & Lorentz rep. & $\SUIIIC$ & $\SUIIL$ & $\UIY$ & $\mathbb{Z}_2$\\
		\midrule
		$\Psi$ & 1 & $\sfrac{1}{2}$ & $(\sfrac{1}{2}, 0)$ & $\mathbf{1}$ & $\mathbf{1}$ & $0$ & $-1$ \\
		$\psi$ & 1 & $\sfrac{1}{2}$ & $(\sfrac{1}{2}, 0)$ & $\mathbf{1}$ & $\mathbf{2}$ & $-1$ & $-1$ \\
		$\psi'$ & 1 & $\sfrac{1}{2}$ & $(\sfrac{1}{2}, 0)$ & $\mathbf{1}$ & $\mathbf{2}$ & $1$ & $-1$ \\
		$\phi_i$ & 2 & $0$ & $(0, 0)$ & $\mathbf{1}$ & $\mathbf{3}$ & $0$ & $-1$ \\
		\bottomrule
	\end{tabular}
\end{table}

Following the notation in previous literature,\cite{Bonnet:2012kz,Restrepo:2013aga}
the model T1-3-B with hypercharge parameter $\alpha = 0$ is defined by
extending the SM with the colour-singlet fields in \cref{tab:model_content}.
The present model therefore combines the $\SUIIL$ triplet scalars $\phi_i$ of zero hypercharge
($\UIY$) with an $\SUIIL$ singlet fermion $\Psi$ and doublet fermions $\psi, \psi'$. They are
all odd under the discrete global symmetry $\mathbb{Z}_2$, while the SM fields are even.
The components of the new fields are given by
\begin{equation}
	\Psi = \Psi^0
	,\quad
	\psi = \begin{pmatrix} \psi^0 \\ \psi^- \end{pmatrix}
	,\quad
	\psi' = \begin{pmatrix} \psi'^+ \\ \psi'^0 \end{pmatrix}
	,\quad
	\phi_i = \begin{pmatrix}
		\frac{1}{\sqrt{2}} \phi_i^0 & \phi_i^+ \\
		\phi_i^- & -\frac{1}{\sqrt{2}} \phi_i^0
	\end{pmatrix},
\end{equation}
where superscripts indicate electric charges. To obtain two non-zero neutrino
mass differences, two generations of scalar triplets are required. Since
the scalar triplets have zero hypercharge, they are treated as real,
$(\phi^0_i)^\dagger = \phi^0_i, (\phi^+_i)^\dagger = \phi^-_i$. $\Psi$ has the same
quantum numbers as a $\mathbb{Z}_2$-odd right-handed neutrino, whereas $\psi$
and $\psi'$ together form a $\mathbb{Z}_2$-odd vector-like lepton doublet, which
makes the model automatically anomaly-free. In principle, all neutral field
components are possible DM candidates. The $\mathbb{Z}_2$ symmetry not only
stabilises the lightest new particle against decay into SM fields, but also
forbids neutrino masses from a tree-level type-I seesaw mechanism.

The most general renormalisable Lagrangian for the model is
\begin{align}
 \mathcal{L} &= \mathcal{L}_{\text{SM}} + \mathcal{L}_{\text{kin}}
 - \frac{1}{2} (M_\phi^2)^{ij} \Tr(\phi_i \phi_j)
 - \lr \frac{1}{2} M_\Psi \Psi \Psi + \text{H.\,c.} \rr
 - \lr M_{\psi\psi'} \psi \psi' + \text{H.\,c.} \rr \nonumber\\
 &\quad
 - (\lambda_2)^{ij} H^\dagger \phi_i \phi_j H
 - (\lambda_3)^{ijkm} \Tr(\phi_i \phi_j \phi_k \phi_m) \nonumber\\
 \label{eq:lagrangian}
 &\quad
 - \lr \lambda_4 (H^\dagger \psi') \Psi + \text{H.\,c.} \rr
 - \lr \lambda_5 (H \psi) \Psi + \text{H.\,c.} \rr
 - \lr (\lambda_6)^{ij} L_i \phi_j \psi' + \text{H.\,c.} \rr,
\end{align}
where $H$ is the SM Higgs field (with vacuum expectation value $v$ and quartic coupling $\lambda$).
The couplings $\lambda_4$ and $\lambda_5$ have the function of Yukawa terms,
which link the fermion singlet and doublets to the SM Higgs boson.
The coupling $\lambda_6$ connects the SM lepton doublet $L$ to the new fields, so
that it will be involved in the process of radiative neutrino mass generation.


\section{Radiative neutrino masses}
\label{sec:radiative_m_nu}

After electroweak symmetry breaking, neutrino masses in our model arise from
a single one-loop diagram. Only the $n_{\text{f}}$ neutral fermion fields $\chi_k$ and
$n_{\text{s}}$ neutral scalar fields $\eta_l$ contribute to mass
generation, whereas the charged fields enter only into the propagator
correction. In our model, the $n_{\text{f}} = 3$ fermions are superpositions of $\SUIIL$
singlets and doublets, while the $n_{\text{s}} = 2$ scalars are superpositions of the
two generations of scalar triplets required for two non-zero neutrino masses.
Evaluating the two-point function in dimensional regularisation and summing
over all contributions (with the neutral fermionic and scalar mixing matrices $U_\chi$ and $O_\eta$) leads to
\begin{equation}
	(M_\nu)_{ij}
	= \frac{1}{32\pi^2} \sum_{l=1}^{n_{\text{s}}} \lambda_6^{im} \lambda_6^{jn} (O_\eta)_{lm} (O_\eta)_{ln}
	\sum_{k=1}^{n_{\text{f}}} {(U_\chi)^*_{k3}}^2 \frac{m_{\chi_k^0}^3}{m_{\eta^0_l}^2 - m_{\chi_k^0}^2} \ln\lr \frac{m_{\chi_k^0}^2}{m_{\eta^0_l}^2} \rr.
	\label{eq:neutrino_masses}
\end{equation}

As evident from \cref{eq:neutrino_masses}, the structure of the mass matrix is
chiefly determined by the number of scalar generations, which must be at least
as large as the desired number of non-zero neutrino masses.
The neutrino mass matrix $M_\nu$ depends explicitly on the couplings $\lambda_6$
and on the masses $m_{\chi^0_k}$ and $m_{\eta^0_l}$ of the neutral fermions
and scalars, while the dependence on the other couplings $\lambda_{1,4,5}$
remains implicit in the mixing matrices. The Casas--Ibarra parametrisation~\cite{Casas:2001sr} then allows to obtain $\lambda_6$ from the experimental
neutrino data, once the other couplings and masses have been fixed.

\section{Dark matter direct detection and lepton flavour violation}
\label{sec:dd_lfv}

\begin{figure}
	\centerline{%
		\includegraphics[width=0.65\textwidth]{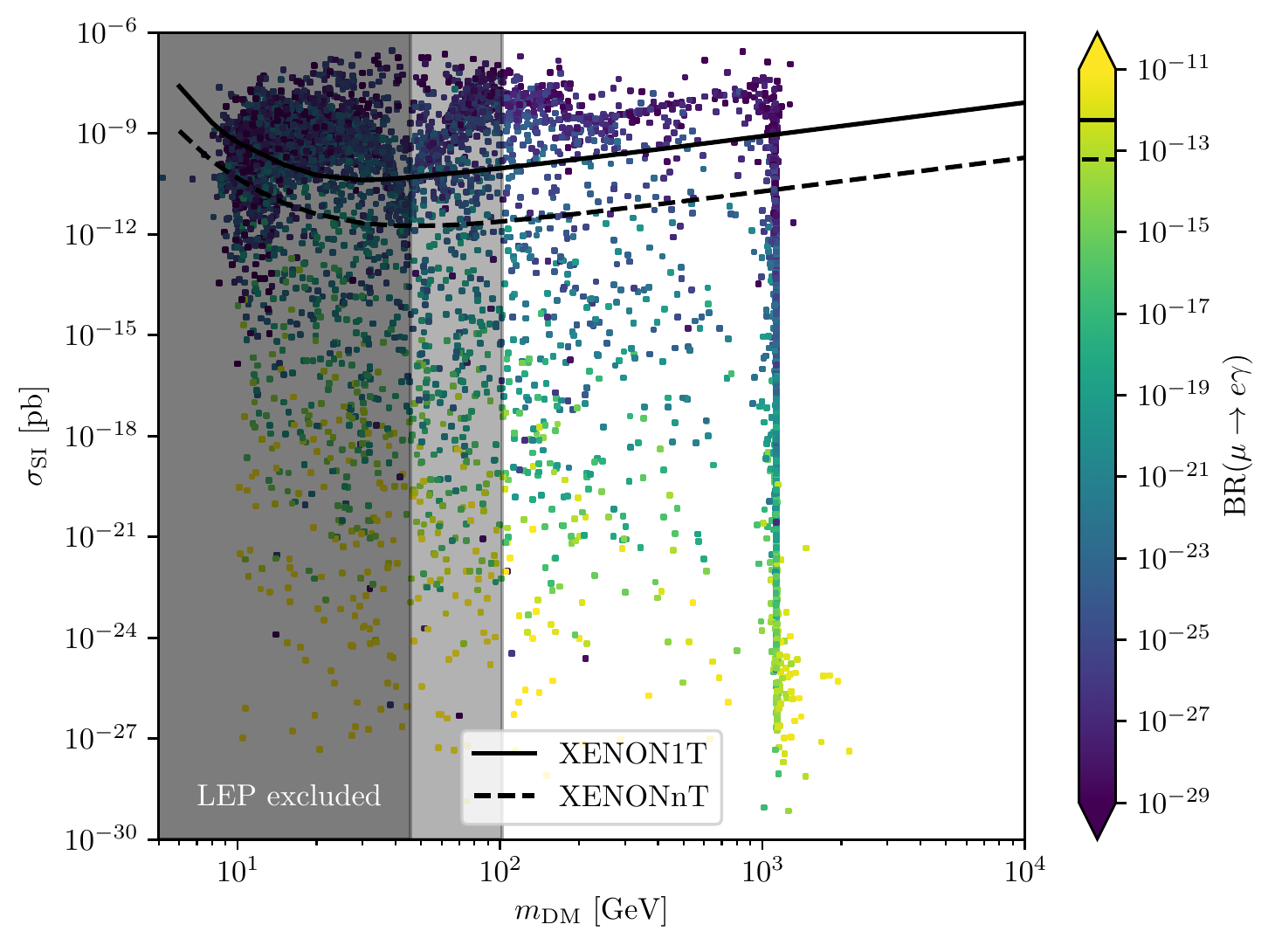}%
		\hspace*{-0.5em}%
		\includegraphics[width=0.65\textwidth]{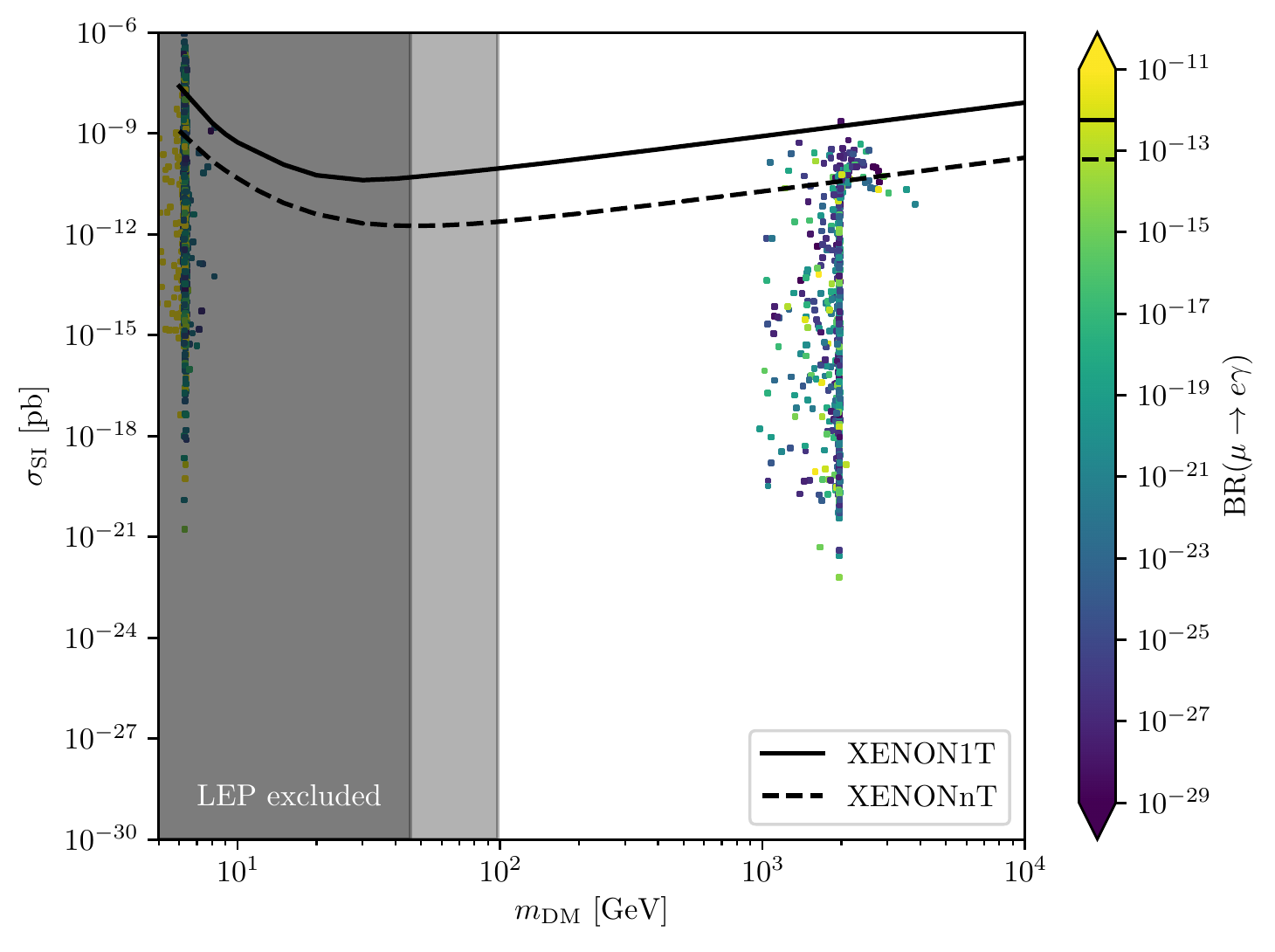}%
	}
	\caption[]{%
		The spin-independent direct detection cross section as a function
		of the DM mass for singlet--doublet fermion DM (left) and triplet scalar DM (right). The colours show the branching
		ratios for the LFV process $\mu \to e \gamma$.
		Also shown are the LEP limits on light neutral and charged particles (shaded areas) as well as current (full lines) and future (dashed lines) exclusion limits for the DM relic density from XENON1T~\cite{Aprile:2018dbl} and XENONnT~\cite{Aprile:2015uzo}, and for $\mu \to e \gamma$~\cite{Adam:2013mnn,Baldini:2013ke}.%
	}
	\label{fig:dm_direct_det_lfv}
\end{figure}

We now connect singlet--doublet fermions with triplet scalars with the aim to not only explain the
observed small neutrino masses as described in the previous section, but also
in order to study the effect of the neutrino mass constraints on the nature,
allowed parameter space, direct detection prospects and LFV properties of the two DM candidates
in this combined model.
We explore the model's parameter space with the help of \texttt{SARAH}~4.13.0~\cite{Staub:2013tta}, calculating the physical
particle spectrum and relevant precision observables with \texttt{SPheno}~4.0.3~\cite{Porod:2011nf} as well as the DM relic density and direct detection cross
sections with \texttt{micrOMEGAs}~4.3.5~\cite{Barducci:2016pcb} using a random parameter scan.

All models resulting from the random scan with the observed neutrino masses and mixings as well as the correct DM relic density
$\OmegacObs h^2 = \num{0.120 +- 0.001}$ and Higgs mass are shown in \cref{fig:dm_direct_det_lfv} as a function of the DM
mass, together with their spin-independent direct detection cross
section and the branching ratio for the usually most sensitive LFV process
$\mu \to e \gamma$.

For fermionic DM, the models accumulating at a DM mass of about \SI{1}{\TeV}
feature mostly doublet fermions, whereas lighter fermionic DM is generally
a superposition of singlet and doublet.
XENON1T excludes most of the
models with small scalar-fermion couplings $\lambda_6$ and therefore also
little LFV. These models are therefore similar to those in the pure
singlet--doublet fermion DM
model. The combination with the scalar sector opens up a considerable
parameter space of leptophilic DM. Interestingly, one observes
a strong complementarity with LFV experiments, which already probe the models
with the smallest spin-independent direct detection cross section.\cite{Adam:2013mnn}

Similarly to the fermionic DM case, LEP constraints already rule out light scalar DM candidates.
As for a pure triplet scalar model, we observe an accumulation of points
around a mass of \SI{2}{\TeV}. Many of these models have only very small couplings
$\lambda_6$ to the fermion sector and thus very little LFV. As $\lambda_1$
increases, so must the DM mass beyond \SI{2}{\TeV} to compensate for the stronger
Higgs annihilation. However, most of these models will soon be probed by
XENONnT, and those that will not can soon be excluded by the process
$\mu \to e \gamma$. While the mass region from \SIrange{1}{2}{\TeV} with leptophilic
fermion DM, that was opened up by coupling the fermion and scalar sectors, was
already excluded by LFV limits (see above), the corresponding models with
scalar DM are still allowed, but will soon be probed by the process
$\mu \to e \gamma$.

\section{Conclusion}
\label{sec:conclusion}

We have combined the singlet--doublet fermion model
with the triplet scalar model in order to explain not only the observed
DM relic density, but also the neutrino masses and mixings, which were
generated radiatively. This model allows in addition for the correct
Higgs boson mass, couplings of natural size, masses in the \si{\TeV} range
and gauge coupling unification.

We found that DM in our model is fermionic up to the \si{\TeV} scale and scalar
beyond. The
scalar--fermion couplings opened the parameter space, so that leptophilic
singlet--doublet fermion DM around \SI{1}{\TeV} became again
viable below the XENON1T
exclusion limit, as did triplet scalar DM between \SIlist{1; 2}{\TeV}. In both
regions, we observed an interesting complementarity between the expectations
for XENONnT and for LFV experiments.

\section*{Acknowledgements}

This work has been supported by the BMBF under contract 05H18PMCC1 and the DFG
through the Research Training Group~2149 \enquote{Strong and weak interactions --
from hadrons to dark matter}. We thank U.\ Oberlack for communication on the
XENONnT expectations and C.\ Yaguna for helpful comments on the manuscript.

\section*{References}

\bibliography{bib}

\begin{thebibliography}{10}

\bibitem{Bonnet:2012kz}
Florian Bonnet, Martin Hirsch, Toshihiko Ota, and Walter Winter.
\newblock {Systematic study of the $d=5$ Weinberg operator at one-loop order}.
\newblock {\em JHEP}, 07:153, 2012.
\newblock arXiv:1204.5862 [hep-ph].

\bibitem{Restrepo:2013aga}
Diego Restrepo, Oscar Zapata, and Carlos~E. Yaguna.
\newblock {Models with radiative neutrino masses and viable dark matter
  candidates}.
\newblock {\em JHEP}, 11:011, 2013.
\newblock arXiv:1308.3655 [hep-ph].

\bibitem{Fiaschi:2018rky}
Juri Fiaschi, Michael Klasen, and Simon May.
\newblock {Singlet-doublet fermion and triplet scalar dark matter with
  radiative neutrino masses}.
\newblock {\em JHEP}, 05:015, 2019.
\newblock arXiv:1812.11133 [hep-ph].

\bibitem{Klasen:2016vgl}
Sonja Esch, Michael Klasen, David~R. Lamprea, and Carlos~E. Yaguna.
\newblock {Lepton flavor violation and scalar dark matter in a radiative model
  of neutrino masses}.
\newblock {\em Eur. Phys. J.}, C78(2):88, 2018.
\newblock arXiv:1602.05137 [hep-ph].

\bibitem{Esch:2018ccs}
Sonja Esch, Michael Klasen, and Carlos~E. Yaguna.
\newblock {A singlet doublet dark matter model with radiative neutrino masses}.
\newblock {\em JHEP}, 10:055, 2018.
\newblock arXiv:1804.03384 [hep-ph].

\bibitem{Hagedorn:2016dze}
Claudia Hagedorn, Tommy Ohlsson, Stella Riad, and Michael~A. Schmidt.
\newblock {Unification of Gauge Couplings in Radiative Neutrino Mass Models}.
\newblock {\em JHEP}, 09:111, 2016.
\newblock arXiv:1605.03986 [hep-ph].

\bibitem{Casas:2001sr}
J.~A. Casas and A.~Ibarra.
\newblock {Oscillating neutrinos and $\mu \to e \gamma$}.
\newblock {\em Nucl. Phys.}, B618:171--204, 2001.
\newblock arXiv:hep-ph/0103065.

\bibitem{Aprile:2018dbl}
E.~Aprile et~al.
\newblock {Dark Matter Search Results from a One Ton-Year Exposure of XENON1T}.
\newblock {\em Phys. Rev. Lett.}, 121(11):111302, 2018.
\newblock arXiv:1805.12562 [astro-ph.CO].

\bibitem{Aprile:2015uzo}
E.~Aprile et~al.
\newblock {Physics reach of the XENON1T dark matter experiment}.
\newblock {\em JCAP}, 1604(04):027, 2016.
\newblock arXiv:1512.07501 [physics.ins-det].

\bibitem{Adam:2013mnn}
J.~Adam et~al.
\newblock {New constraint on the existence of the $\mu^+ \to e^+\gamma$ decay}.
\newblock {\em Phys. Rev. Lett.}, 110:201801, 2013.
\newblock arXiv:1303.0754 [hep-ex].

\bibitem{Baldini:2013ke}
A.~M. Baldini et~al.
\newblock {MEG Upgrade Proposal}.
\newblock 2013.
\newblock arXiv:1301.7225 [physics.ins-det].

\bibitem{Staub:2013tta}
Florian Staub.
\newblock {SARAH 4: A tool for (not only SUSY) model builders}.
\newblock {\em Comput. Phys. Commun.}, 185:1773--1790, 2014.
\newblock arXiv:1309.7223 [hep-ph].

\bibitem{Porod:2011nf}
W.~Porod and F.~Staub.
\newblock {SPheno 3.1: Extensions including flavour, CP-phases and models
  beyond the MSSM}.
\newblock {\em Comput. Phys. Commun.}, 183:2458--2469, 2012.
\newblock arXiv:1104.1573 [hep-ph].

\bibitem{Barducci:2016pcb}
D.~Barducci et~al.
\newblock {Collider limits on new physics within micrOMEGAs\_4.3}.
\newblock {\em Comput. Phys. Commun.}, 222:327--338, 2018.
\newblock arXiv:1606.03834 [hep-ph].

\end{thebibliography}

\end{document}